\documentclass[a4paper,12pt]{spieman}  
\usepackage{amsmath,amsfonts,amssymb}
\usepackage{graphicx}
\usepackage{setspace}
\usepackage{tocloft}
\usepackage{tablefootnote}
\usepackage{multirow}
\usepackage{multicol}

\title{Characterization of the SKA1-Low prototype station Aperture Array Verification System 2}

\author[a,*]{Giulia Macario}
\author[b]{Giuseppe Pupillo} 
\author[b]{Gianni Bernardi}
\author[a]{Pietro Bolli}
\author[a]{Paola Di Ninni}
\author[a]{Giovanni Comoretto}
\author[b]{Andrea Mattana}
\author[b]{Jader Monari}
\author[b]{Federico Perini}
\author[b]{Marco Schiaffino}
\author[c]{Marcin Sokolowski}
\author[c]{Randall Wayth}
\author[c]{Jess Broderick}
\author[d]{Mark Waterson}
\author[d]{Maria Grazia Labate}
\author[e]{Riccardo Chiello}
\author[f]{Alessio Magro}
\author[c]{Tom Booler}
\author[c]{Andrew Mcphail}
\author[c]{Dave Minchin} 
\author[c]{Raunaq Bhushan}

\affil[a]{Istituto Nazionale di Astrofisica (INAF), Osservatorio Astrofisico di Arcetri, Largo Enrico Fermi 5,  Firenze, Italy, 50125}
\affil[b]{Istituto Nazionale di Astrofisica (INAF), Istituto di Radioastronomia, Via Piero Gobetti 101, Bologna, Italy, 40129}
\affil[c]{International Centre for Radio Astronomy Research (ICRAR), Curtin University, Perth,  Australia, GPO Box U1987, 6845 }
\affil[d]{SKA Observatory, Jodrell Bank, Lower Withington, Macclesfield, UK}   
\affil[e]{University of Oxford, Denys Wilkinson Building, Oxford, UK}   
\affil[f]{Institute of Space Sciences and Astronomy, University of Malta, Msida, Malta}    

\cftpagenumbersoff{figure}
\cftpagenumbersoff{table} 
\begin{document} 
\maketitle

\begin{abstract}
The low frequency component of the Square Kilometre Array (SKA1-Low) will be an aperture phased array located at the Murchison Radio-astronomy Observatory (MRO) site in Western Australia. It will be composed of 512 stations, each of them consisting of 256 log-periodic dual polarized antennas, and will operate in the low frequency range (50 MHz - 350 MHz) of the SKA bandwidth. 
The Aperture Array Verification System 2 (AAVS2), operational since late 2019, is the last full-size engineering prototype station deployed at the MRO site before the start of the SKA1-Low construction phase. The aim of this paper is to characterize the station performance through commissioning observations at six different frequencies (55, 70, 110, 160, 230 and 320 MHz) collected during its first year of activities. We describe the calibration procedure, present the resulting all-sky images and their analysis, and discuss the station calibratability and system stability. Using the difference imaging method, we also derive estimates of the SKA1-Low sensitivity for the same frequencies, and compare them to those obtained through electromagnetic simulations across the entire telescope bandwidth, finding good agreement (within 13\%). Moreover, our estimates exceed the SKA1-Low requirements at all the considered frequencies, by up to a factor of $\sim$2.3. Our results are very promising and allow an initial validation of the AAVS2 prototype station performance, which is an important step towards the upcoming SKA-Low telescope construction and science.
\end{abstract}


{\noindent \footnotesize\textbf{*} Macario, G.,  \linkable{giulia.macario@inaf.it} }

\begin{spacing}{1}   

\section{Introduction}
\label{sect:intro}  

The Square Kilometre Array (SKA) will be the world's largest and most-sensitive radio-telescope ever built, covering the huge bandwidth from 50 MHz to 15.3 GHz. It will bring a revolution in astronomy and astrophysics in the upcoming decades\cite{braun2019}~. When completed, it will consist of two telescopes: the SKA-Mid  for the highest frequency end of the bandwidth (above $350$ MHz, in the Karoo region of South Africa) and the SKA-Low for the lowest (below $350$ MHz, in the Murchison Desert of Western Australia).

The SKA-Low, to be built in the upcoming years at the Murchison Radio-astronomy Observatory (MRO) site, will operate with the unprecedented sensitivity and resolution in the frequency range from 50 MHz to 350 MHz. In its current design for Phase 1, SKA1-Low will be an aperture phased-array telescope consisting of 131072 fixed dual polarization log periodic antennas, arranged into 512 stations of about 40 m diameter each containing 256 antennas\cite{braun2019}~, with highly flexible digital beam-forming driven by advanced electronics. Almost 50\% of the stations will be located in a very dense core of $\approx$1 km diameter, while the remaining stations will be distributed along three quasi spiral arms, with a maximum baseline of 65 km\cite{dewdney2020}~.

The deployment of prototype stations is a crucial step in the engineering design and development process before the construction of large and complex systems using new and/or unproven technology such as SKA-Low. Since 2016, various full-size prototype SKA-Low stations have been deployed and tested at the MRO (e.g. AAVS1\cite{benthem2021}~; EDA1\cite{wayth2017}). 
In this paper we will focus on the latest prototype station built at MRO, the Aperture Array Verification System 2 (hereafter AAVS2\cite{davidson2020,vanes2020}~),  deployed in 2019 and currently operational along with the comparator system Engineering Development Array 2  (EDA2, see Wayth et al.\cite{wayth2021}, submitted to this journal, for details). 
In the following sections, after a brief description of AAVS2 (Section \ref{sec:backgr}), we present a first characterization of the array performance based on the most relevant results obtained through AAVS2 observations since its first light along with the simulations (Sections \ref{sec:obs} and \ref{sec:sens}). 

\begin{figure}[htbp]
\begin{center}
\includegraphics[width=1.0\textwidth]{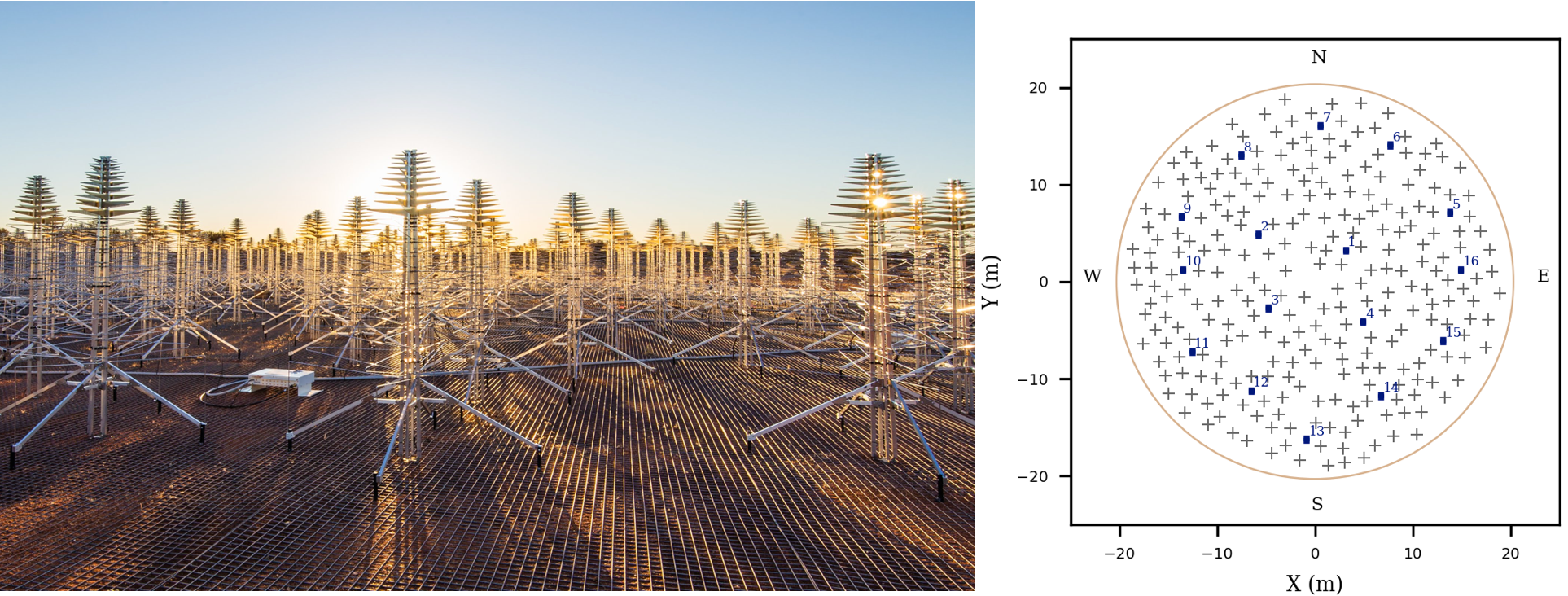}
\end{center}
\caption 
{ \label{fig:aavs2_Lay}
 \textit{Left panel}: AAVS2 picture showing some of the SKALA4.1 aluminum log-periodic antennas. Credits: INAF/ICRAR \textit{Right panel:} AAVS2 array layout, showing positions of the 256 antennas (black crosses), pseudo-randomly distributed within $\sim$40 m diameter area (yellow circle). The blue rectangles indicate the position of the 16 SMART boxes.} 
\end{figure} 

\section{Background}
\label{sec:backgr}

The AAVS2 is the most recent full-sized SKA-Low station prototype, based on the the experience gained from the commissioning results of its predecessor AAVS1 \cite{benthem2021,vanes2020}~. 
It consists of 256 SKALA4.1 log periodic dual polarization antennas with Low Noise Amplifiers (LNAs) optimized to meet the required performances in the frequency bandwidth 50-350 MHz\cite{bollirep2020}~. Antennas are pseudo-randomly distributed over a circular area of about 40 m diameter, with a layout chosen to improve the overall performance of the station\cite{vanes2020}~. The layout is basically identical to that of EDA2 and the predecessor stations\cite{wayth2021,benthem2021}~, although the diameter of AAVS2 has been increased by $\sim$10\% to host the larger SKALA4.1 antennas (the maximum distance between antenna centers is $\approx$38 m\cite{sokolPasa2021})~.
A picture of the AAVS2 station is shown in Fig.\ref{fig:aavs2_Lay}, left panel), along with the station layout, showing the positions of the antennas with respect to the center of the array (right panel). Antennas are fixed to a wire ground mesh aligned with local cardinal directions (Fig.\ref{fig:aavs2_Lay}, left panel), so that antenna dipoles are oriented North-South and East-West\cite{wayth2021}~. 
Coaxial cables connect antennas to 16 \textquotedblleft SMART\textquotedblright\, boxes (one for every 16 antennas, Fig.\ref{fig:aavs2_Lay}, right panel), each one converting radio frequency to optical signals. 
The signals, after being aggregated in a nearby Field Node Distribution Hub (FNDH), are then transmitted through a $\sim$5.5 Km optical fiber connection to the shielded room at the MRO control building, where the analog receivers and the acquisition and digital elaboration systems based on Field Programmable Gate Array (FPGA) are located. In the control room, optical signals are connected to 16 Tile Processing Modules (TPMs), where different data outputs are produced. 
For a complete description of the system, signal chain, monitoring and control software, observing modes and data processing, we refer the reader to previous works that provide more details (see \cite{wayth2017,benthem2021,vanes2020,wayth2021,sokol2021}).

\section{Observations and data processing}
\label{sec:obs}

Observations with the AAVS2 station as a stand-alone interferometer were carried out at coarse channel central frequencies $\nu_c =$ 54.7, 70.3, 110.2, 159.4, 229.8 and 320.3~MHz to commission and test the array. Throughout the paper we will refer to these central frequencies as 55, 70, 110, 160, 230 and 320~MHz, for simplicity.
Data were taken in April 2020 apart for the 54.7~MHz frequency that was observed on February, 19$^{\rm th}$ 2021, all of them spanning a time interval of at least 22~hours  (Table~\ref{tab:obs}). Observations consisted of a series of 0.28~s snapshots (separated into two 0.14~s integrations)
, using a single $\approx$925.926~kHz wide coarse channel.

Visibilities were generated from the complex voltages of individual antennas using a software correlator\cite{clark2011,sokolPasa2021}~, yielding 32 channels, each $\approx$28.935~kHz-wide. They were stored using the UVFITS format\cite{greisen2017}~. 
Static delays (due to cable length differences) were corrected before channelisation, such that all signals are time-aligned to within a $\approx$1.08 $\mu$s sample. The remaining residual differential delays between antennas (that are also stable on timescales of month\cite{benthem2021}) were corrected by applying phase calibration solutions from daily solar calibration scans to each coarse channel observation (as done for AAVS1\cite{benthem2021} and EDA2\cite{wayth2021}). 

\begin{figure}[t]
\begin{center}
\includegraphics[width=0.6\textwidth]{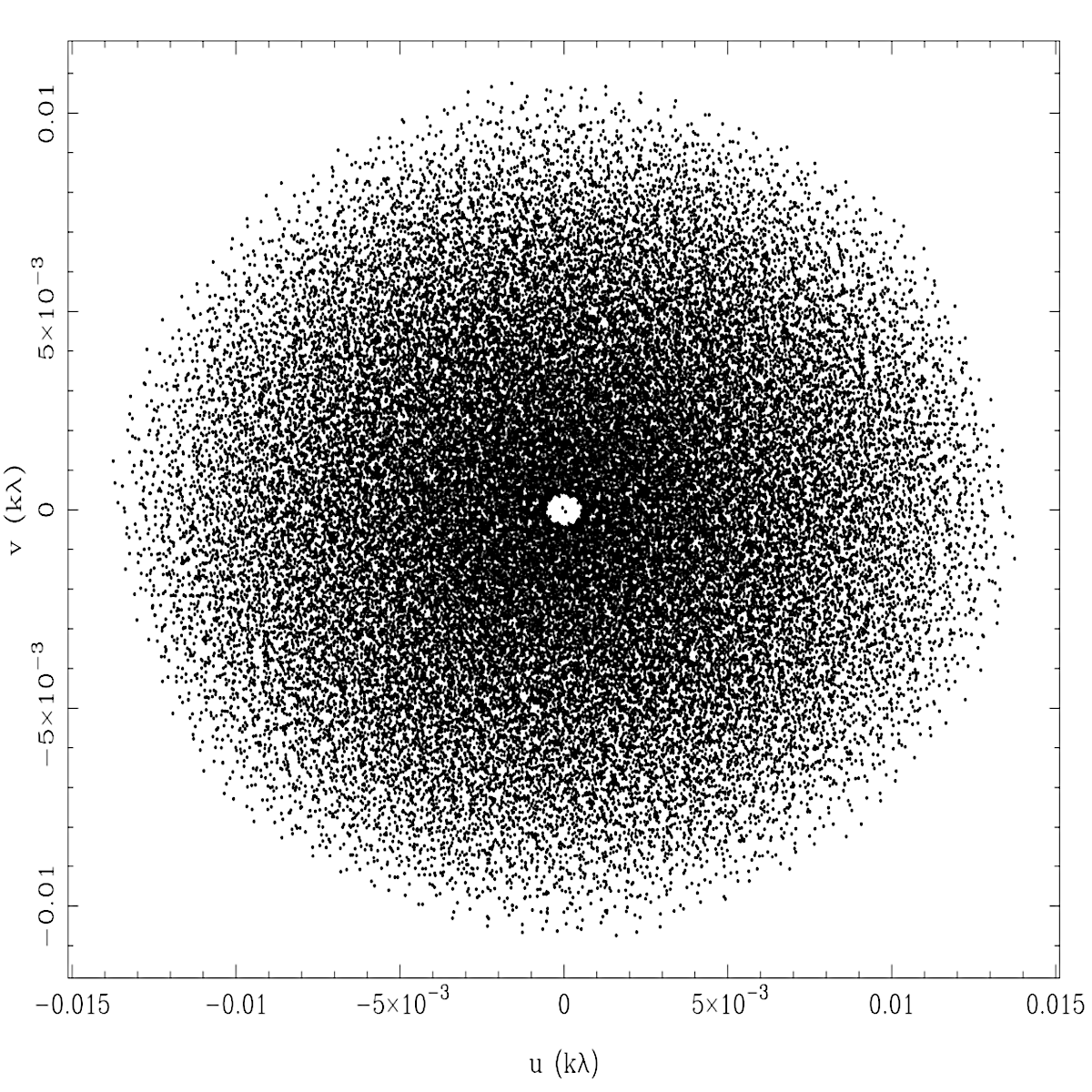}
\end{center}
\caption 
{ \label{fig:uvcov}
AAVS2 uv-coverage of the 110 MHz observation of 2020/04/22, for the snapshot of solar transit (single timestep, XX polarization) and integration time $\simeq0.14$ s.
}
\end{figure}

Data reduction was carried out with the Miriad package\cite{sault1995}~. Six antennas were permanently flagged in the April 2020 data and 12 in data from 
February 2021. Five edge channels (the first 3 and the last 2) were flagged too, reducing the bandwidth to $\approx$780 kHz. Fig.~\ref{fig:uvcov} shows an example of the $uv$ coverage of the array at 110~MHz, for the first timestep of a single snapshot with $\approx$0.14~s integration time.

\begin{table}[htbp]
\caption{Summary of AAVS2 observations used in this work.} 
\label{tab:obs}
\begin{center}       
\begin{tabular}{|c|c|c|c|} 
\hline
\rule[-1ex]{0pt}{3.5ex}  $\nu_c$  &  Start time, UT &  LST coverage & UT at Sun transit  \\
\rule[-1ex]{0pt}{3.5ex}    (MHz) &  (yyyy/mm/dd, hh:mm:ss) & (hours) &  (yyyy/mm/dd, hh:mm:ss) \\
\hline
\rule[-1ex]{0pt}{3.5ex}  $54.7$  & $2021/02/19$, $03:03:01$ & $\simeq22$ &  $2021/02/19$, $04:27:58$ \\
\hline
\rule[-1ex]{0pt}{3.5ex}  $70.3$ & $2020/04/17$, $08:59:25$ & $\simeq24$ &  $2020/04/18$, $04:13:44$ \\
\hline
\rule[-1ex]{0pt}{3.5ex}  $110.2$  & $2020/04/21$, $10:19:31$ & $\simeq24$ &  $2020/04/22$, $04:13:53$ \\
\hline
\rule[-1ex]{0pt}{3.5ex}  $159.4$ & $2020/04/07$, $16:26:58$ & $\simeq24$ &  $2020/04/08$, $04:16:33$ \\
\hline
\rule[-1ex]{0pt}{3.5ex}  $229.8$  & $2020/04/19$, $03:52:11$ & $\simeq22$ &  $2020/04/19$, $04:12:10$ \\
\hline
\rule[-1ex]{0pt}{3.5ex}  $320.3$  & $2020/04/22$, $11:09:11$ & $\simeq24$ &  $2020/04/23$, $04:13:34$ \\
\hline
\end{tabular}
\end{center}
\end{table} 

\begin{table}[ht]
\caption{Power-law model flux densities of the Sun at the central frequencies $\nu_c$ of the coarse channels (last two columns), derived from the Benz measurements \cite{benz2009} (first and second column) with spectral indexes (fourth column) computed in the closest frequency ranges (third  column).} 
\label{tab:benz}
\begin{center}       
\begin{tabular}{|c|c||c|c|c|c|} 
\hline
\rule[-1ex]{0pt}{3.5ex}  $\nu_i$\cite{benz2009}  & $S_i$\cite{benz2009} & $\Delta\nu$& $\alpha$ &  $\nu_c$  & $S_{\nu_c}$\\
\rule[-1ex]{0pt}{3.5ex}   (MHz) & (Jy) & (MHz) &  &  (MHz) & (Jy) \\
\hline
\hline
\rule[-1ex]{0pt}{3.5ex}  $50$  & $5400$ & $50-100$ & $2.15$ & $54.7$ & $6550$  \\
\rule[-1ex]{0pt}{3.5ex}  & & &  & $70.3$ & $11200$  \\
\hline
\rule[-1ex]{0pt}{3.5ex}  $100$  & $24000$ & $100-150$ & $1.86$ &  $110.2$ & $28700$   \\
\hline
\rule[-1ex]{0pt}{3.5ex}  $150$ & $51000$ & $150-200$ &  $1.61$ & $159.4$ & $56200$   \\
\hline
\rule[-1ex]{0pt}{3.5ex}  $200$  & $81000$ & $200-300$ &  $1.50$ & $229.8$ & $99800$  \\
\hline
\rule[-1ex]{0pt}{3.5ex}  $300$  & $149000$ & $300-400$ & $1.31$ &  $320.3$ & $162000$  \\
\hline
\end{tabular}
\end{center}
\end{table}

The Sun is unresolved at all frequencies by the longest baselines of the array and was therefore used as point-like calibration source (similar to previous works\cite{benthem2021,vanes2020,sokol2021})~. During our observations there was no sign of solar activity either (see \linkable{https://www.solarmonitor.org/index.php}). 
The low frequency solar spectrum is well measured down to $30$~MHz \cite{benz2009}~. In the 50-350 MHz SKA1-Low frequency range, we approximated it with a power-law model with spectral index $\alpha$ changing with the frequency interval:
\begin{equation}
\label{eq:sunmodel}
S_{\nu_c} = S_i \left( \frac{\nu_c}{\nu_i} \right)^{\alpha}, 
\end{equation}
where $\nu_i$ and $S_i$ are measurements from observations of the quiet Sun\cite{benz2009}~, and  $\alpha$ is computed between the two $\nu_i$ values nearest to each central frequency $\nu_c$ (see third column of Table~\ref{tab:benz}). 
Calibration proceeded in the same way for each frequency: for the snaphots when the Sun was closer to its transit (last column of Table~\ref{tab:obs}) 
visibilities were rotated to the Sun position (by means of the task UVEDIT) and the complex bandpass was derived for each time-step independently using the MFCAL task with a 0.14~s solution interval. In order to minimize the contribution from the Galactic emission, baselines $\leq2\lambda$ at frequencies $\leq$ 70 MHz and $\leq5\lambda$ at the other higher frequencies were excluded from the calibration.
Calibration solutions obtained from the snapshots corresponding to the solar transit were transferred (through the GPCOPY task) to all the other snapshots.  
Sun-based calibration failed at 55~MHz, where the solar emission is too faint with respect to the Galactic background emission. To set the absolute flux density scale at this frequency was thus applied the procedure described in Section \ref{sec:55cal}. 

The antenna primary beam response was not used directly in the calibration model.  We rather fixed the absolute flux density scale by multiplying each flux density and noise measurement of our analysis (see Sections \ref{sec:55cal},\ref{sec:fluxstab}, \ref{sec:sens}) by the primary beam response corresponding to the Sun position in the snapshot observation used for calibration, normalized to zenith. 
We assumed that all the antennas have the same primary beam, the average embedded element pattern (hereafter EEP)\cite{davidson2020,bolli2021}~.

Visibilities were Fourier transformed into zenith phase-centered all--sky images, using the MIRIAD task INVERT, with natural weighting and a pixel size of 21~arcmin - corresponding to 4--23 pixels across the synthesized beam, depending upon the observing frequencies. For every snapshot across the entire observing window, each polarization was imaged separately, generating $XX$ and $YY$ images at each frequency.
Each image covered the entire hemisphere visible at a specific time, with the synthesized beam ranging from $\approx$8.0$^{\circ}$ 
to $\approx$1.4$^{\circ}$ with increasing frequency (see Table \ref{tab:images}).
Dirty images were deconvolved using the more appropriate algorithm (between H\"ogbom, Clark and Steer, i.e. Miriad CLEAN mode set to default\cite{miriad}~a, with a maximum of 200 iterations. 
The quality of the obtained images remains typically good at all the analyzed frequencies across 24 hours, indicating that the system is relatively stable over long timescales.

\begin{figure}[htbp]
\begin{center}
\includegraphics[width=1.\textwidth]{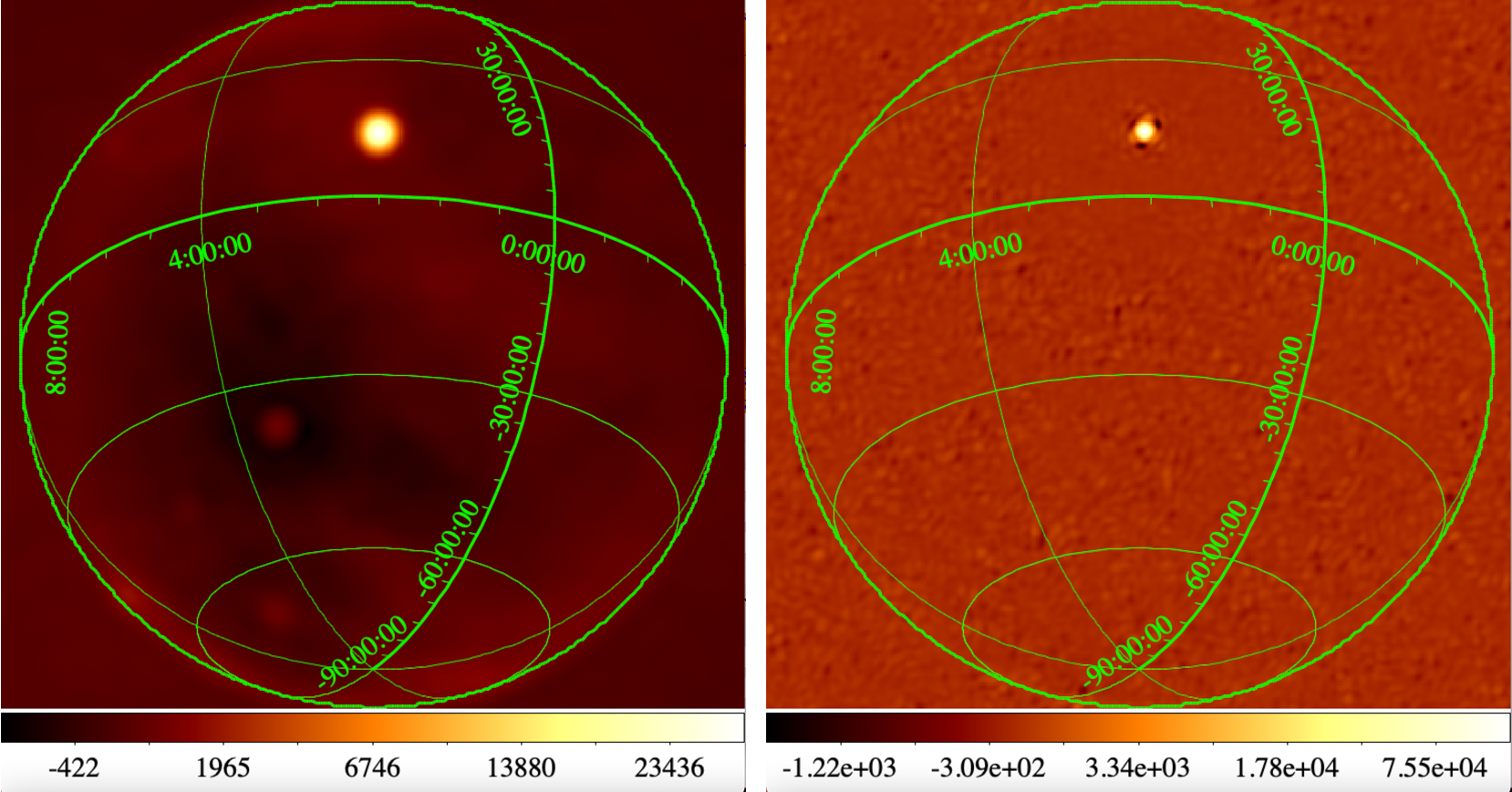}
\end{center}
\caption 
{ \label{fig:suntransit110320}
All-sky, $XX$ polarization images at solar transit at 110~MHz (left panel) and 320~MHz (right panel). Units are Jy~beam$^{-1}$.}
\end{figure}

\begin{figure}[htbp]
\begin{center}
\includegraphics[width=1.0\textwidth]{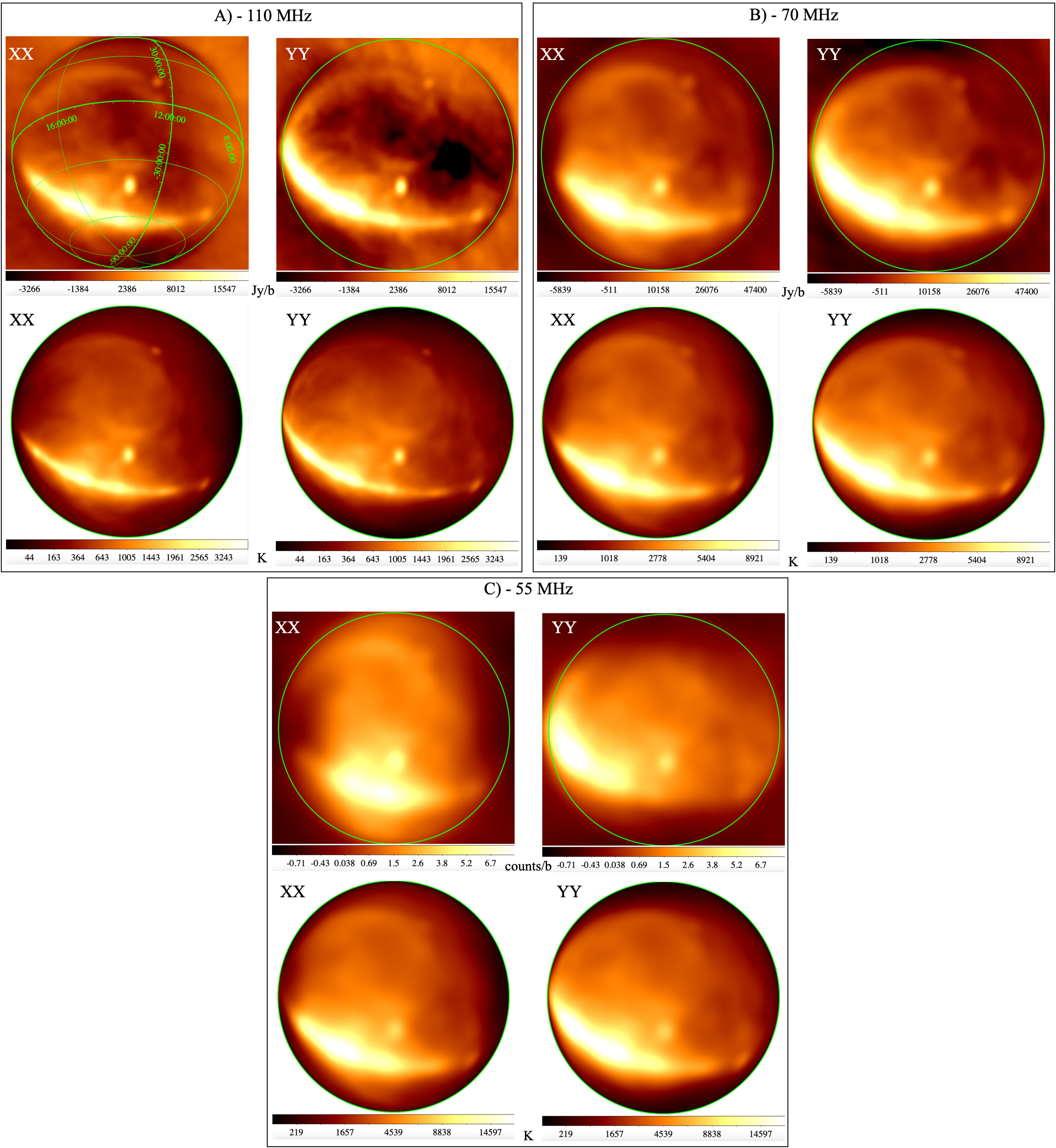}
\end{center}
\caption 
{ \label{fig:images}
All-sky, snapshot images at the transit of Centaurus~A, for $XX$  and $YY$ polarizations at 110~MHz (panel A), 70~MHz (panel B) and 55~MHz (panel C). In each panel, the observed sky brightness (top frames) in Jy beam$^{-1}$ (counts/beam for un-calibrated images at 55 MHz) is compared to the corresponding $T_{sky}$ simulations (bottom frames), in K. Square root color scales are in the following ranges (for data and simulation, respectively): $-6500$ - $60000$ Jy beam$^{-1}$, $4$ - $4000$ K at 110 MHz;  $-3500$ - $20000$ Jy beam$^{-1}$ , $30$ - $11000$ K at 70 MHz; $-0.8$ - $8.5$ counts beam$^{-1}$ , $40$ - $18000$ K at 55 MHz. The green circle indicates the horizon.
}
\end{figure}

Fig.~\ref{fig:suntransit110320} shows images at 110 and 320~MHz corresponding to the snapshot at solar transit (see last column of Table \ref{tab:obs}), 
obtained by averaging the two timesteps. The Sun is clearly the dominant source in the sky, showing that the assumption of a point-like calibration model is appropriate.
Fig.~\ref{fig:images} shows examples of snapshot images at 110, 70 and 55~MHz, taken when Centaurus~A, clearly visible at all frequencies, is approximately at transit ( $\approx 12$~hours after solar culmination, i.e. calibration snapshot). Moreover, at the AAVS2 resolution, the Galactic plane is the brightest visible feature, consistently detected across frequencies.
Fig.~\ref{fig:images} also shows a comparison between simulated sky temperature images and actual observations. The observable sky model at each frequency, i.e. the map of the sky brightness temperature above the horizon at a given time from the AAVS2 location, was derived from the global sky model (GSM) \cite{oliveiracosta08}, with contribution from the Sun emission added \cite{benz2009}~. The all sky simulated maps were generated using the PyGSM code (see \linkable{https://github.com/telegraphic/PyGSM}). For a better comparison with the actual observations, the sky model maps were smoothed to match the AAVS2 angular resolution at each frequency (see Table \ref{tab:images}). 
We note a qualitative good agreement between the observed and the simulated structures (upper and lower panels of Figure \ref{fig:images}, respectively), where Centaurus A is at the expected position. The Galactic plane has a more pronounced morphological difference, particularly at 55 MHz for the $XX$ polarization and close to the horizon (panels C, left side). This discrepancy may be due to mutual coupling from the antennas\cite{davidson2020,bolli2021}, which is more prominent near the horizon and is not accounted for in our simulations. We also note that our simulated images do not include the station $uv$ coverage and this can also cause slight differences between observed and simulated images, although a more quantitative comparison is left for the future.

\begin{table}[ht]
\caption{Average synthesized beams of AAVS2 all-sky images, for the analyzed frequencies (see Section \ref{sec:obs} and Figure \ref{fig:images})} 
\label{tab:images}
\begin{center}       
\begin{tabular}{|c|c|} 
\hline
\rule[-1ex]{0pt}{3.5ex}  $\nu$  & $\theta$, P.A. \\
\rule[-1ex]{0pt}{3.5ex}   (MHz) & ($^{\circ}\times^{\circ}$,$^{\circ}$) \\
\hline
\rule[-1ex]{0pt}{3.5ex}  $55$  & $8.0\times 7.9$, $9.0$ \\
\hline
\rule[-1ex]{0pt}{3.5ex}  $70$  & $6.3 \times 6.1$, $-31.0$ \\
\hline
\rule[-1ex]{0pt}{3.5ex}  $110$  & $4.0 \times 3.9$, $-31.1$   \\
\hline
\rule[-1ex]{0pt}{3.5ex}  $160$ & $2.8 \times 2.7$, $-31.1$ \\
\hline
\rule[-1ex]{0pt}{3.5ex}  $230$  & $1.9 \times 1.9$, $-31.0$ \\
\hline
\rule[-1ex]{0pt}{3.5ex}  $320$  &$1.4 \times 1.3$, $-31.1$   \\
\hline
\end{tabular}
\end{center}
\end{table}

\subsection{Calibration at 55~MHz}
\label{sec:55cal}

As mentioned in the previous section, the Sun is not sufficiently bright to serve as a  calibration source at 55~MHz; we therefore adopted the following strategy.
All-sky snapshot images were produced from visibilities after only delays were applied. 
Receiver gains are approximately equalized to have the same wide-band power at the TPM input for all antennas\cite{comoretto2018} and this first order equalization, together with antenna based delays, already allow to obtain  good images (Fig.~\ref{fig:images}, Section \ref{sec:obs}).

\begin{figure}[htbp]
\begin{center}
\includegraphics[width=0.8\textwidth]{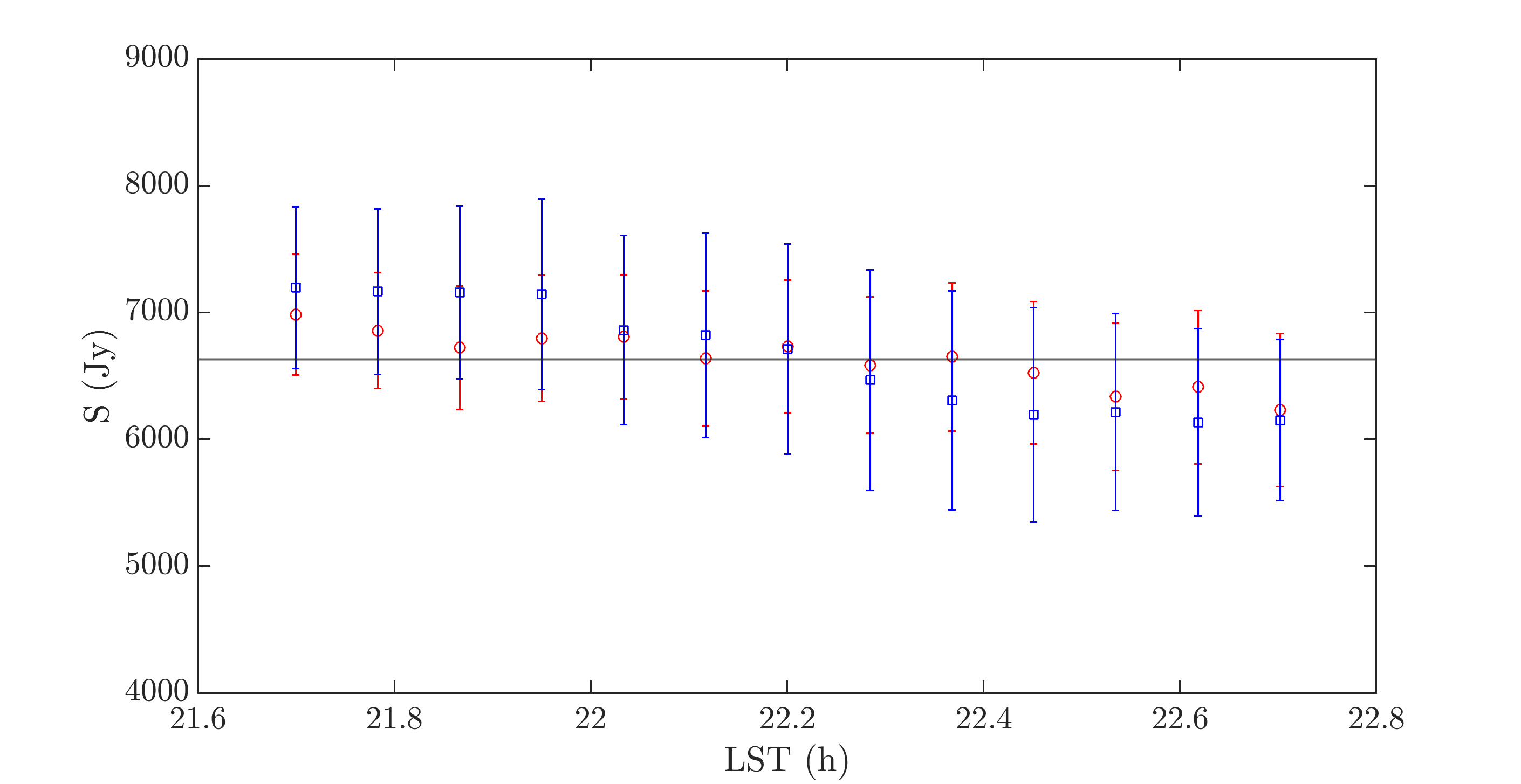}
\end{center}
\caption 
{ \label{fig:sun_55_LST}
Sun flux densities (in Jy) from un-calibrated snapshot images at 55 MHz ($\approx$1 h around Sun transit) with  1$\sigma$ error bars, after primary beam correction and a posteriori absolute flux scale calibration (Section \ref{sec:55cal}), plotted as a function of LST (in h). Red and blue circles correspond to XX and YY polarization data, respectively. Solid line is the Sun reference flux density value (see \cite{benz2009} and Table \ref{tab:benz}).
}
\end{figure} 

The peak brightness values of the Sun and the corresponding uncertainties was extracted for the snapshots in the $21.7 < {\rm LST} < 22.7$~h range ($\approx 1$~h around transit) using the IMFIT task, and corrected for the normalized mean EEP\cite{davidson2020,bolli2021} in the direction of the Sun.  
We found the apparent flux density of the Sun to be sufficiently constant in time (see Fig.~\ref{fig:sun_55_LST}), showing that the simple power equalization does a reasonably good job in calibrating the visibility amplitudes. 

We then calculated the ratio between the expected flux density of the Sun at 55~MHz (Table \ref{tab:benz}) and the apparent flux density, averaged over the snapshots and  corrected for the element primary beam per polarization. This scaling factors were used to bring our measurements at this frequency on the appropriate absolute flux density scale (Sections \ref{sec:fluxstab} and \ref{sect:diffim}). Results are shown in Fig.~\ref{fig:sun_55_LST}, where corrected solar flux densities with 1$\sigma$ error bars for each linear polarization are plotted against LST, and compared to the Sun reference value at 55 MHz. 
The same procedure applied to Centaurus A in $\approx$1 hour around its transit (LST range $\simeq$12.8-13.9~h) provided correction factors within $\leq$10 \%  those derived from the Sun.

\section{Performances}
\subsection{Consistency checks on radio sources flux densities}
\label{sec:fluxstab}

To evaluate the time stability of calibration and assess the quality of the absolute flux scales, we performed the analysis of the flux densities of selected radio sources in the field (other than the Sun), as a function of time, and their consistency across the frequencies. 
Among the brightest \textquotedblleft A-team\textquotedblright\, radio sources, we have selected those with an elevation $\geq$ 60$^{\circ}$ at their culmination.  This criterion allows to analyze the fluxes $\approx$1 hour across the source transit, avoiding the sensitivity losses due to the poor SKALA4.1 antenna response at lower elevations (which is related to the natural drop off in the antenna design trade-off\cite{bolli2020}). Only Centaurus A, Fornax A and Pictor A met this selection criterion. 
We produced lightly cleaned all-sky images of snapshots in the selected LST ranges, with a similar procedure as described in Section \ref{sec:obs}, but averaging the two timesteps and excluding also in the imaging the same baselines discarded in the calibration. 
The signal-to-noise ratios of Centaurus A and Fornax A are $\gtrapprox$~4 - 5 at most of the frequencies, ensuring a good quality of the measured fluxes. 
Pictor A is too weak to be detected by these AAVS2 observations at frequencies $\leq$70 MHz (see Section \ref{sec:obs}). 

\begin{figure}[ht]
\begin{center}
\begin{tabular}{c}
\includegraphics[width=0.9\textwidth]{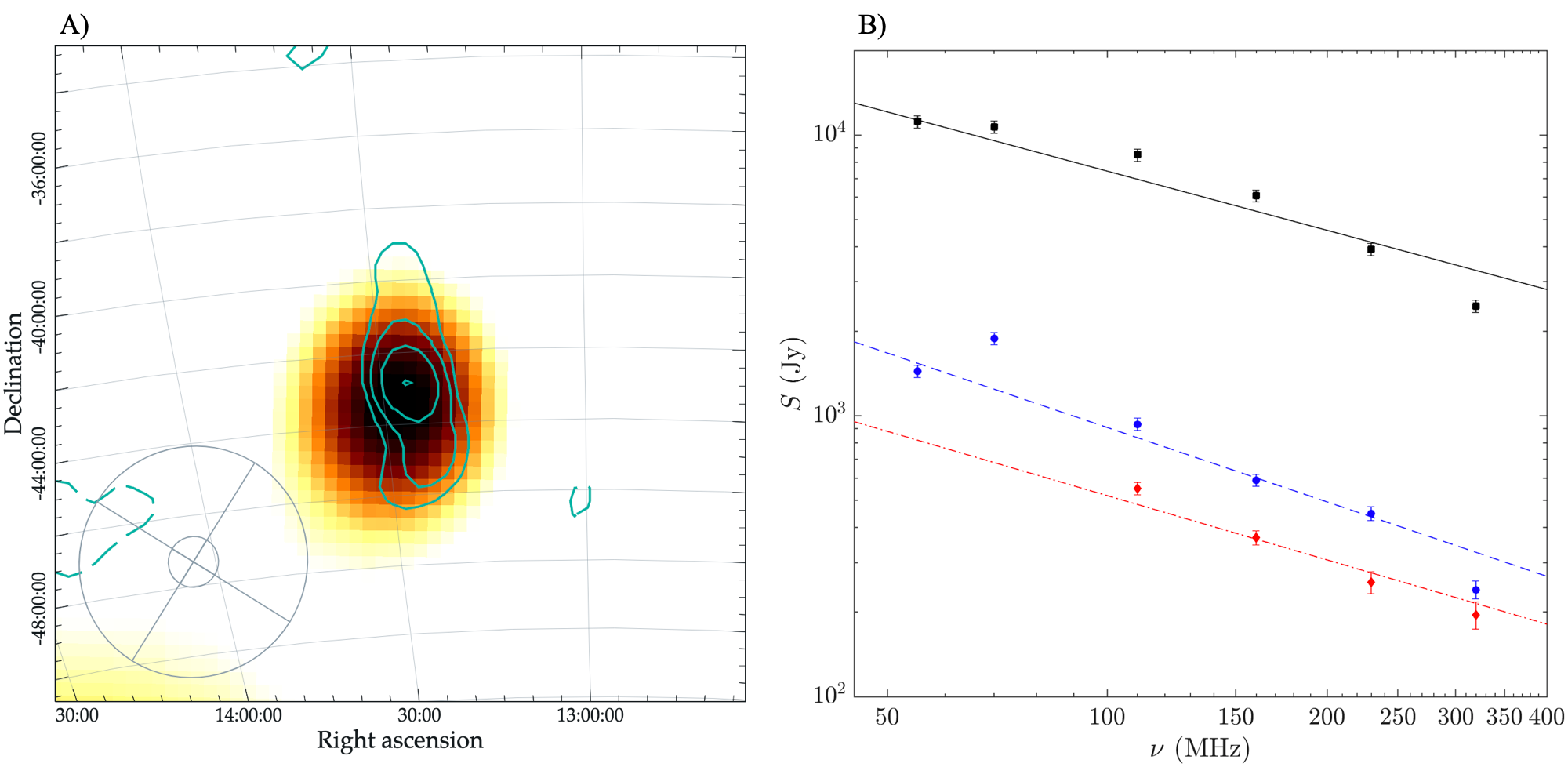}
\end{tabular}
\end{center}
\caption 
{ \label{fig:spectrum}
\textit{Panel A}: 70 MHz image around Centaurus A, with the 320 MHz contours overlaid. Both images are from XX polarization, and include all baselines. The local $\sigma_{rms}$ is $\simeq$1.3$\times$10$^{4}$ Jy beam$^{-1}$ at 70 MHz and $\simeq$100 Jy beam$^{-1}$ at 320 MHz. Color scale (linear) ranges from 2$\times$10$^4$ to 3$\times$10$^4$ Jy beam$^{-1}$; cyan contours are spaced by a factor of 2, starting from $\pm$2.5$\sigma_{rms}$ (dashed are negative). The beam sizes are shown by the gray circles (see Table \ref{tab:images}). \textit{Panel B}: Integrated spectrum of Centaurus A (black squares), Fornax A (blue dots) and Pictor A (red diamonds). The lines are the power-law spectra from literature, for a comparison: Centaurus A (black solid line, $\alpha=-0.70$\cite{mcKin13}), Fornax A (blue dashed line, $\alpha = -0.88 $\cite{bernardi13}), Pictor A (red dotted line, $\alpha = -0.76 $\cite{jacobs2013})~. }
\end{figure}

\begin{figure}[ht!]
\begin{center}
\includegraphics[width=1.0\textwidth]{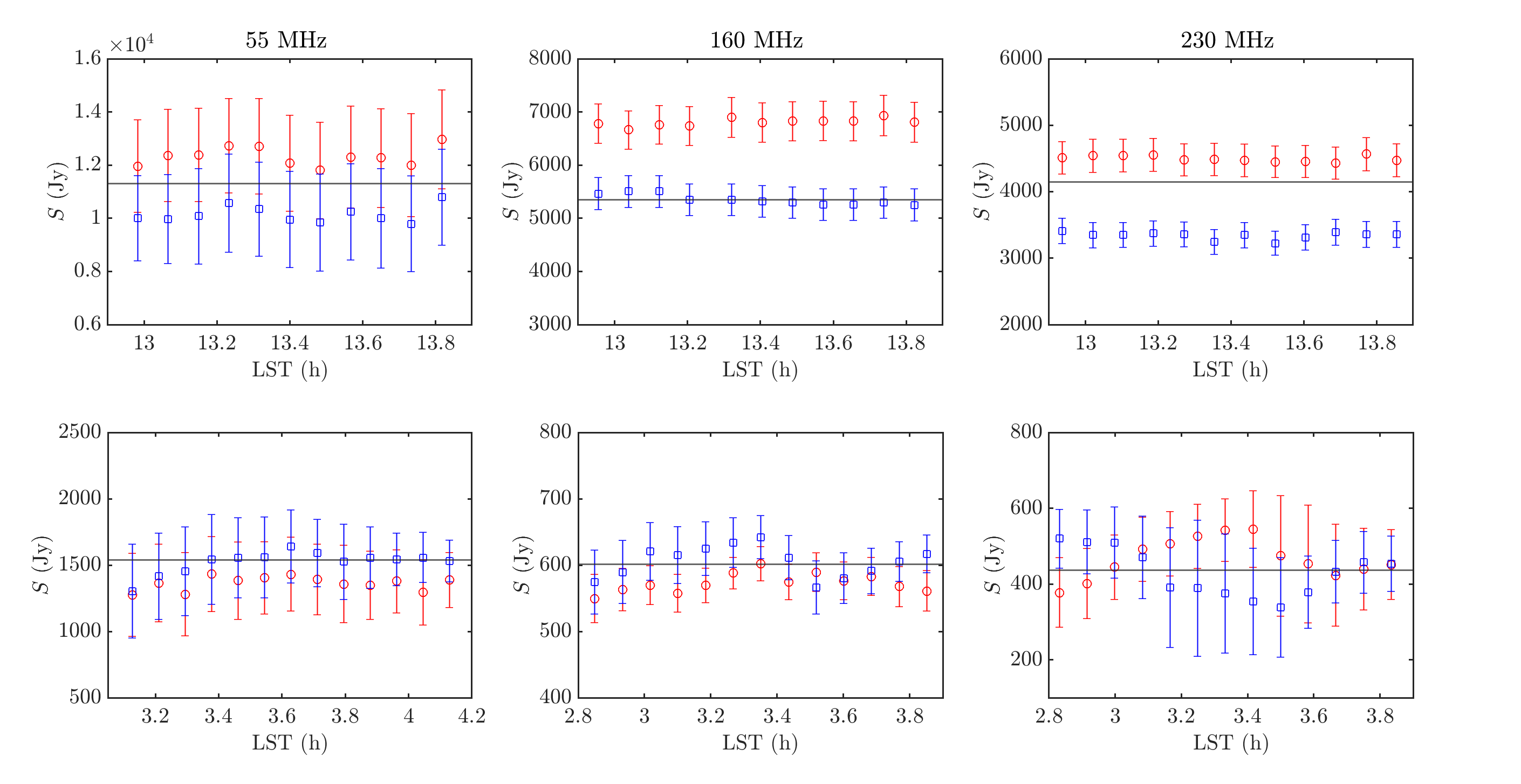}
\end{center}
\caption 
{ \label{fig:fluxLST}
Centaurus A (\textit{Upper plots}) and Fornax A (\textit{Lower plots}) flux densities (in Jy) with 1$\sigma$ error bars as a function of LST (in h), $\approx$1 h around their transit, for three selected frequencies (55, 160 and 230 MHx, from left to right). 
Red circles and blue squares are the measurements from XX and YY polarization images, respectively. The black lines are the corresponding reference values, extrapolated from McKinley et al. 2013\cite{mcKin13} and Bernardi et al. 2013\cite{bernardi13}~. }
\end{figure}

As Fornax A and Pictor A are unresolved at all the frequencies, their peak flux densities were extracted using the Miriad task IMFIT. To improve the accuracy of these measurements, we used a two component fit: a Gaussian with the size of the point-spread-function and an initial offset estimate, in a square region around each  source adequately chosen depending upon the frequency. 
The reference flux density values of Fornax A at each frequency were obtained by extrapolating the value at 189 MHz to the observing frequencies using a single power-law spectrum of spectral index $-0.88$ (Bernardi et al. 2013\cite{bernardi13}). Similarly, the reference values of Pictor A were obtained by extrapolating the measurement at 150 MHz with a single power-law spectrum of spectral index $-0.76$ (Jacobs et al. 2013\cite{jacobs2013}).

Centaurus A is resolved at frequencies $\geq$110 MHz. A zoom around this source at 70 and 320 MHz (XX polarization) is shown in Fig. \ref{fig:spectrum}, panel A). For a proper comparison with the reference values (extrapolated from the value at 118 MHz using a power law of spectral index $-0.70$, given in McKinley et al. 2013 \cite{mcKin13}), we thus obtained its integrated flux densities at those frequencies using the AEGEAN source finder tool (Hancock et al. 2012\cite{hank2012} and 2018 \cite{hank2018}), within islands covering the source extended emission (down to  3$\sigma_{rms}$ level in each analyzed image). As AEGEAN does not provide the uncertainties on the island flux densities (see \cite{hank2012,hank2018}), these were estimated as $\sigma_S = \sigma_{rms}\times\sqrt{N_{\text{beams}}}$, with $\sigma_{rms}$ being the noise level around the source, and $N_{\text{beams}}$ the number of beams crossing source (i.e. the island integration area). 
Peak flux densities and their uncertainties at frequencies $\leq$70 MHz were obtained in the same way as done for Fornax A and Pictor A, both unresolved at these frequencies. 
All the flux measurements and their uncertainties were finally corrected for the primary beam, taking into account the antenna response both in the direction of the Sun (calibrator) and of each radio source (excluding measurements at 55 MHz, already corrected for the primary beam, see Section \ref{sec:55cal}). The average EEP\cite{davidson2020,bolli2021} per frequency and linear polarization was used in this correction.

Primary beam corrected fluxes (in Jy) as a function of the LST (in hours) within $\sim$1 hour across the radio source transit are shown in Fig. \ref{fig:fluxLST}, for three of the observed frequencies (55, 160 and 230 MHz, from left to right). The upper plots refer to Centaurus A, the lower plots to Fornax A. Measurements of both linear polarizations are shown (red circles and blue squares for XX and YY, respectively), and compared to the reference flux values (black lines) derived as described above (see \cite{bernardi13,mcKin13}).

The integrated spectra of Centaurus A (black squares), Fornax A (blue dots) and Pictor A (red diamonds) are shown in Fig. \ref{fig:spectrum}, panel B). Each point in the spectrum (with $1\sigma_{tot}$ error bars) is the average of all the corresponding measurements in the considered LST range and between the two polarizations. The total uncertainty is computed as: $\sigma_{tot}=\sqrt{(S\times\sigma_{amp})^2 + (\sigma_S)^2}$, with $\sigma_{amp}$ being the amplitude errors on the calibrator, conservatively assumed to be $\sim$5\% (see e.g. \cite{scaifeHeald2012,bernardi2009}~), and $\sigma_S$ the uncertainties on the averaged flux density measurements. 

This initial analysis shows that, even with a first order calibration procedure: 1) the measured radio sources flux densities exhibit relatively small variations across the selected LST intervals, of the order of 5-10\% for Centaurus A and 5-20\% for Fornax A at most of the frequencies (Figure \ref{fig:fluxLST}); 2) they are generally consistent with the reference values (Figures \ref{fig:fluxLST} and \ref{fig:spectrum}); 3) the measured spectra of Pictor A, Fornax A and Centaurus A across the SKA1-Low bandwidth follow reasonably  well the expected power laws extrapolated from literature (Figure \ref{fig:spectrum}).

\subsection{Sensitivity}
\label{sec:sens}

In this section we describe the analysis performed to estimate the SKA1-Low sensitivity through AAVS2 observations, using the difference imaging technique. This method has been successfully applied to estimate the sensitivity of AAVS1 prototype station\cite{benthem2021} and for a preliminary verification of AAVS2  sensitivity\cite{sokol2021} (using mostly the same data as in this paper). 

\subsubsection{Difference imaging}
\label{sect:diffim}

We used the same observations described in Section \ref{sec:obs} (see Table \ref{tab:obs}) to derive sensitivities at 55, 70, 110, 160, 230 and 320 MHz. 
The difference imaging method is based on the generation and analysis of the differences between each pair of close-in-time images, over which the sky and the calibration do not appreciably change. The image differences should thus, ideally, include just noise (all the astrophysical radio sources and calibration artifacts cancel out, see  \cite{benthem2021,sokol2021})~. The difference between the two time-step images (Section \ref{sec:obs}) for each snapshot and each polarization were produced through the MATHS task. The rms of the noise in all the image differences was measured through the IMHIST task within three square boxes of increasing sizes (41$\times$41, 61$\times$61, 81$\times$81 pixels, respectively) centered around zenith. 

\subsubsection{Sensitivity measurements across LST}

For each coarse channel, visual inspection of the difference images produced showed that they are mostly noise-like and free of residual emission. 
The few images with significant remaining artifacts from the difference process (due to e.g. interference present in one of the two time step images) have been excluded from the sensitivity analysis. The three values of the rms of the noise extracted from the zenith-centered square boxes in each difference image were averaged, obtaining one measurement per linear polarization: $\sigma_{p}$, (with $p$ indicating the linear polarization $XX$, $YY$). 
As in the calibration we did not correct the model flux densities of the Sun for the antenna response (Section \ref{sec:obs}), we applied the primary beam corrections to the $\sigma_{p}$  measurements, by multiplying those for the antenna response in the direction of the Sun (normalized to zenith). We used the average EEP\cite{davidson2020,bolli2021} corresponding to each central frequency and linear polarization. At 55 MHz, these measurements  (already corrected for the primary beam) were re--scaled to the flux density scale as described in Section \ref{sec:55cal}. 
The corrected $\sigma_{p}$ values were then converted into the station System Equivalent Flux Density according to equation:

\begin{equation}
\label{eq:sefd}
SEFD_{s,p} \simeq \frac{\sigma_p}{\sqrt{2}}\eta \sqrt{t_i B}\frac{N}{256} \;\;\;\text{Jy} \, ,
\end{equation}

where $p$ indicates the linear polarization ($XX$, $YY$), $N$ is the number of used station antennas in the observations (250 for April 2020 data and $238$ for February 2021 data, see Section \ref{sec:obs}), $t_i\approx$0.14s is the integration time of each input image (assumed to be the same for all datasets), $B\approx0.78$ MHz is the effective bandwidth of each dataset (after the edge channels flagging, Section \ref{sec:obs}), $\eta$ is the system efficiency (assumed to be equal to 1). The factor $\sqrt{2}$ results from the assumption that the two input images used to make the image difference have identical Gaussian noise characteristics, thus the rms of the image difference is a factor $\sqrt{2}$ the actual noise of a single snapshot image. Note moreover that the total intensity is defined as $I =  (I_{XX} + I_{YY})/2$.  
Equation \ref{eq:sefd} was then used to compute the measured station sensitivity: 

\begin{equation}
\label{eq:stationsens}
s_{s,p} \equiv  \frac{A_{eff}}{T_{sys}}=10^{26}\frac{2k}{SEFD_{s,p}} \;\;\;\text{m}^2\text{K}^{-1} \, ,
\end{equation}

where k is the Boltzmann constant ($k=1.38 10^{-23}$ m$^{2}$ kg s$^{-2}$ K$^{-1}$), $A_{eff}$ is the effective area of the station and $T_{sys}$ is the system temperature. The measured $SEFD$ of the entire SKA1-Low array is thus simply $SEFD_{s,p}$ divided by 512 (the total number of SKA1-Low stations): 

\begin{equation}
\label{eq:sefdska}
SEFD_{SKA}=\frac{SEFD_{s,p}}{512} \;\;\;\text{m}^2\text{K}^{-1} \, ,
\end{equation}

The measured SKA1-Low sensitivity was thus given by: 

\begin{equation}
\label{eq:skasens}
s_{SKA,p} \equiv  \frac{A_{eff}}{T_{sys}}=10^{26}\frac{2k}{SEFD_{SKA,p}} \;\;\;\text{m}^2\text{K}^{-1} \, ,
\end{equation}

\begin{figure}[htbp]
\begin{center}
\includegraphics[width=0.75\textwidth]{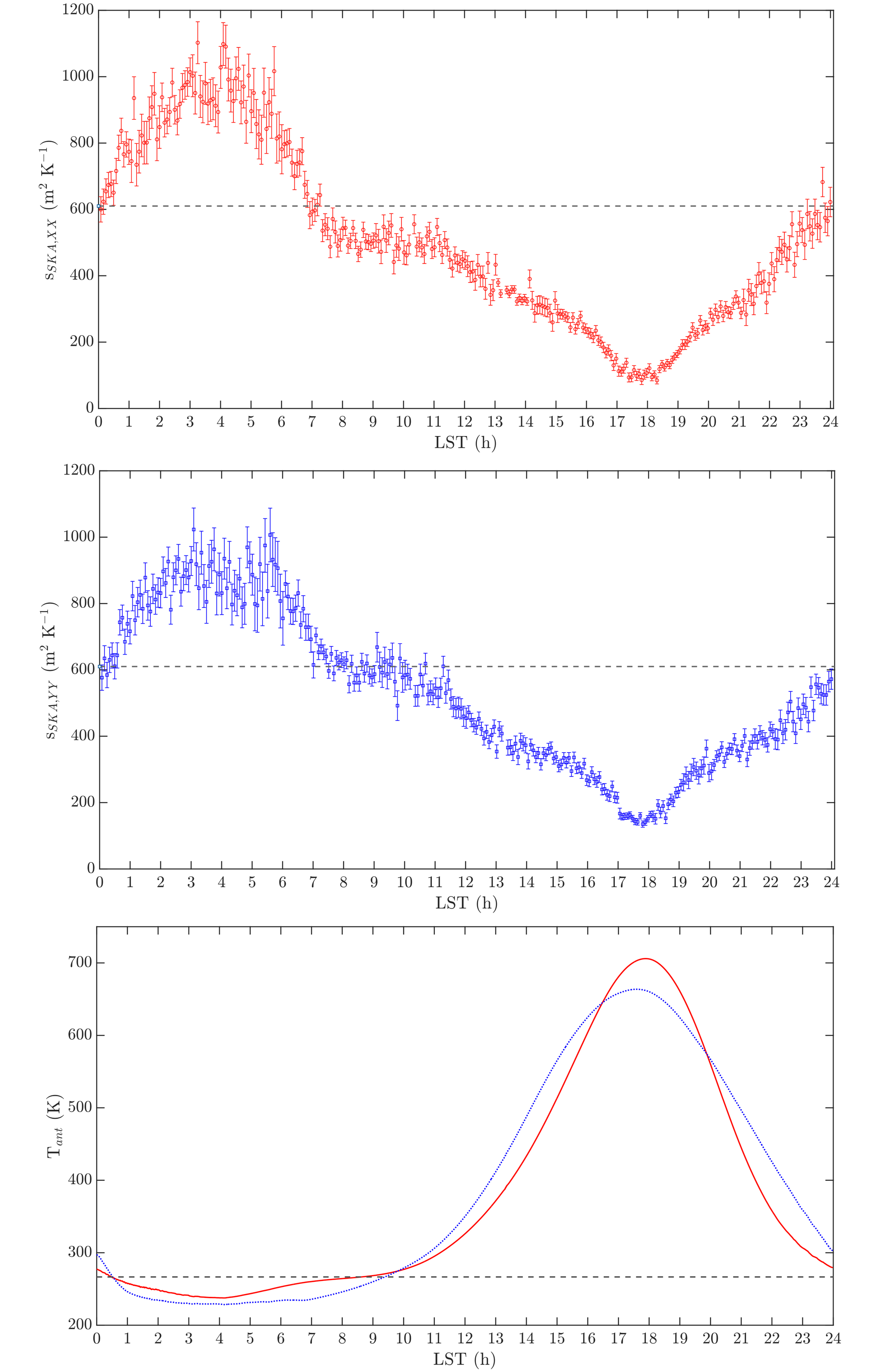}
\end{center}
\caption 
{ \label{fig:sensCurve}
\textit{Upper and central panels}: 2020/04/07-08, 160 MHz $s_{SKA,p}$ sensitivity (in m$^2$ K$^{-1}$) as a function of LST (in h), compared to the SKA1-Low sensitivity requirement (dashed line, \cite{caiazzo}); red circles and blue squares are estimates for $XX$ and $YY$ polarizations, respectively, with 1$\sigma$ error bars. \textit{Lower panel}: Corresponding $T_{ant}$ (in K) simulation as a function of LST for the same date and time range; red solid and blue dotted lines refer to XX and YY polarization, respectively. The dashed line is the T$_{sky}$ value used to compute the SKA1-Low sensitivity requirements\cite{caiazzo}~. 
}
\end{figure}

The estimated SKA1-Low sensitivities curves as a function of LST across the full length of the  observation (22 to 24 hours) were derived at all the analyzed frequencies. 
An example of these curves at 160 MHz is reported in Fig. \ref{fig:sensCurve} (upper and mid panels), for the two linear polarizations. The corresponding SKA1-Low sensitivity requirement\cite{caiazzo} is also shown with the dashed line. The uncertainties (1$\sigma$ error bars in the plots) are estimated every 1 hour interval as the standard deviation of the individual sensitivity measurements, after subtraction of the baseline polynomial fit of the full 24 hrs data. As expected, the highest sensitivity values are reached when the Galactic Plane is below the horizon and the Sun is close to its transit (LST $\sim3-4$ h at 160 MHz, in Fig. \ref{fig:sensCurve}, upper and central panels); the sensitivity then decreases as the Galactic Plane emission rises and moves across the field of view, reaching its minimum value when it transits above the array (LST $\sim18$ h at 160 MHz, in Fig. \ref{fig:sensCurve}, upper and central panels). This is evident in the plot of antenna temperature as a function of LST (see Fig. \ref{fig:sensCurve}, \textit{lower panel}), estimated as:

\begin{equation}
\label{eq:tant}
T_{ant} (\nu) =  \frac{\int_{4\pi} P_{p}(\nu,\theta,\phi)T(\nu,\theta,\phi)d\Omega}{\int_{4\pi} P_{p}(\nu,\theta,\phi)d\Omega} ,
\end{equation}

where $P_{p}(\nu,\theta,\phi)$ is the average EEP (per polarization,\cite{davidson2020,bolli2021}), $T(\nu,\theta,\phi)$ is the sky brightness temperature at frequency $\nu$ and pointing direction ($\theta,\phi$) simulated as described in Section \ref{sec:obs}.
The sensitivity estimates for the two linear polarizations, averaged over the entire LST ranges, are consistent each other within $\lesssim$15\% at all the analyzed frequencies except at 70 MHz, where $XX$ values are a factor of $\sim$2 higher with respect to $YY$ estimates. This discrepancy might be related to mutual coupling effects (see \cite{davidson2020,bolli2021}), however it needs to be further investigated.

We note that, even if our calibration procedure takes into account the fact that the calibrator source is not at zenith (Section \ref{sec:obs}), our sensitivity measurements most likely underestimate the actual SKA1-Low zenith sensitivity, that would be reachable through an ideal calibration leading to thermal noise.  

For frequencies $\geq 70$ MHz, self calibration with the Sun was also applied, by selecting sub-sets of snapshots in the LST ranges corresponding to Sun elevations $\geq +45^{\circ}$, to avoid calibration inaccuracies related to the \textquotedblleft 
naturally\textquotedblright\, degraded sensitivity of the antenna at low elevations\cite{bolli2020}~. The sensitivities derived through self calibration are consistent with those obtained through calibration using a single snapshot at the Sun transit (in the common LST intervals) for all the frequencies $\leq$160 MHz, suggesting a good system calibration stability over several hours (see also \cite{vanes2020}). However, at frequencies $\geq230$ MHz we found offsets between the two sensitivity estimates, with single snapshot calibration providing systematically lower values with respect to self calibration. As self calibration removes the time dependencies of the system, we expect such sensitivity estimates to be more accurate. For this reason, the sensitivity measurements at 230 and 320 MHz obtained with single snapshot calibration across the full length of the observations were consistently re-scaled (as detailed in \cite{pupillo20}).

\subsubsection{Comparison with sensitivity simulations and requirements}
\label{sec:sensimul}

In this section we present a comparison of our experimental estimates of the SKA1-Low sensitivity with the requirements and the sensitivity derived from electromagnetic (EM) simulations across the entire bandwidth ($50-350$ MHz). As the requirements are specified for a uniform sky temperature model\cite{caiazzo}, i.e. without considering its variability over the 24 hours and its spatial distribution (see Fig.\ref{fig:sensCurve}, lower panel), for this comparison we use our mean estimates in the LST range $0-8$~h (see Table \ref{tab:sensit}). 
In this LST range, the actual T$_{ant}(\nu)$ during the AAVS2 observations is closer to the T$_{sky}(\nu)$ assumed for computing the requirement sensitivity values (see Section \ref{sec:sens} and Fig. \ref{fig:sensCurve}, lower panel). Moreover, this range corresponds to a \textquotedblleft cold\textquotedblright\, sky patch in which the SKA1-Low sensitivity requirements are defined\cite{caiazzo}. 

\begin{figure}[h!]
\begin{center}
\includegraphics[width=0.87\textwidth]{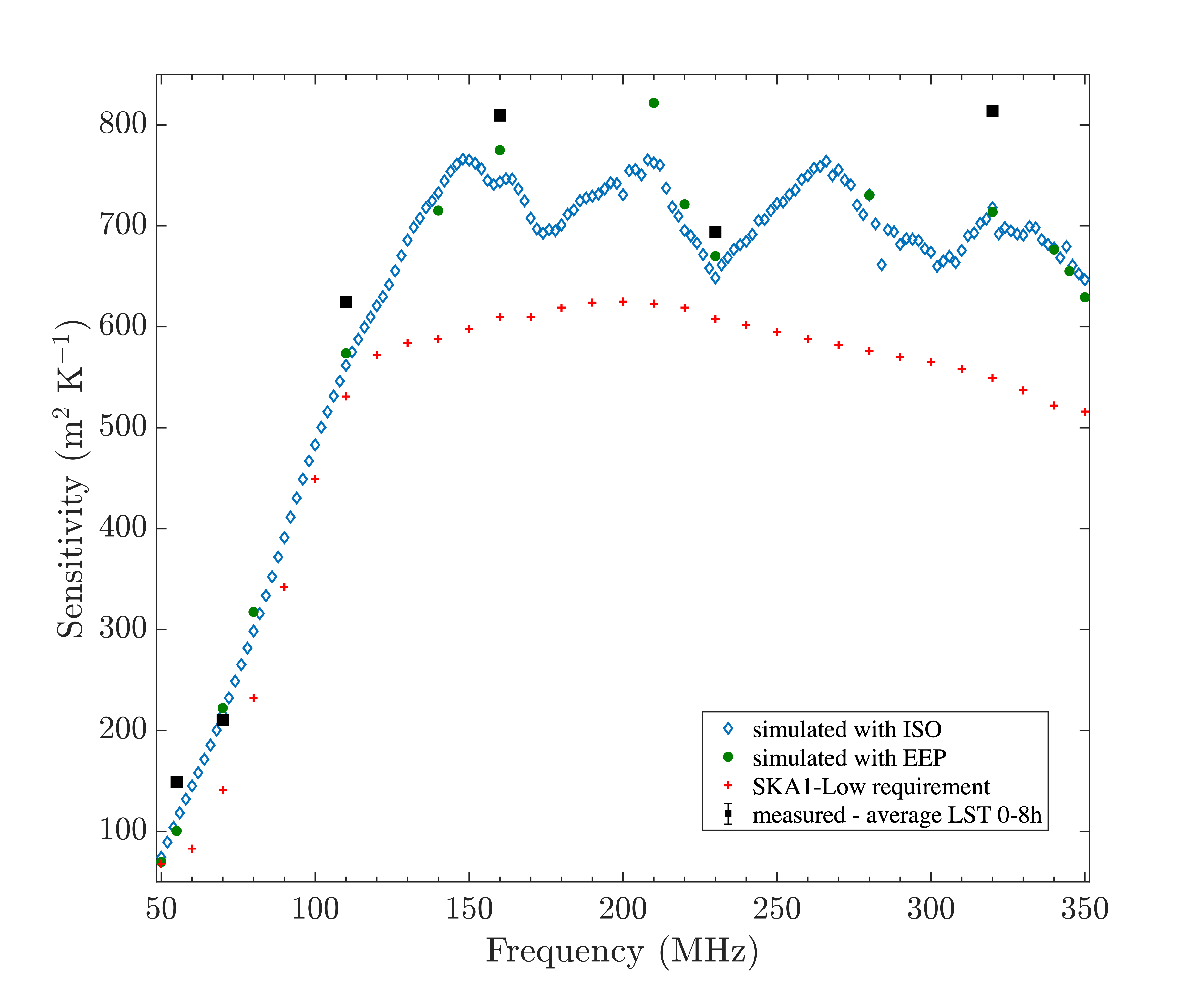}
\end{center}
\caption 
{ \label{fig:sensBW}
Zenith sensitivity over the SKA1-Low bandwidth (averages of XX and YY polarizations). Black squares are the mean sensitivity estimates in LST $0-8$ h (Table \ref{tab:sensit}), with 1$\sigma$ error bars within the symbol size (Section \ref{sec:sens}). Blue diamonds and green circles are the simulated sensitivities computed using the isolated SKALA4.1 antenna patterns and the EEP, respectively. Red crosses are the SKA1-Low requirements\cite{caiazzo} (see Table \ref{tab:sensit}).
}
\end{figure} 

Results are shown in Fig. \ref{fig:sensBW}: black squares are the SKA1-Low mean sensitivity estimates (in m$^2$K$^{-1}$) as a function of frequency (in MHz; see Table \ref{tab:sensit}). Error bars are included within the symbol size, and correspond to  1$\sigma$ standard deviations of the averaged values, i.e. the square root of the variance of the average over the LST interval $0-8$ hours and two polarizations. Note that these do not include the uncertainties on the absolute flux scale, assumed to be $\sim$5\% (see also Section \ref{sec:fluxstab}). 
These estimates are compared to EM simulations of sensitivity at zenith  for one station\cite{bolli2021}, and here re-scaled for the whole SKA1-Low: blue diamonds result from the sensitivity computed from the isolated SKALA4.1 pattern (therefore without mutual coupling effects), while green circles are from the EEP, and both of them are averages of the XX and YY polarizations. Overall, the agreement between the two simulated sensitivities at zenith is very good, meaning that the mutual coupling does not deteriorate the sensitivity (see \cite{davidson2020,bolli2021}). 
The simulation of sensitivity from the isolated patterns has been executed for frequencies in the range 50 MHz to 350 MHz, with a step of 2 MHz. The simulation from the EEP has been performed at 50, 55, 70, 80, 110, 140, 160, 210, 220, 230, 280, 320, 340, 345, 350 MHz.  The requirements are shown with the red crosses. 
As stated in Section \ref{sec:sens}, our measurements are more likely underestimates of the actual SKA1-Low zenith sensitivity, that would be reachable in ideal, thermal noise limited images. Moreover our estimates are averages in LST $0-8$ h, thus not exactly comparable with these simulations. We find however an overall good agreement for most of the analyzed frequencies, with measured averages (black squares of Figure \ref{fig:sensBW}) generally higher than the EEP simulated sensitivity (green circles of Figure \ref{fig:sensBW}), with differences that range between $\sim$3.5\% (at 230 MHz) and $\sim$13\% (at 320 MHz). 
Finally, all our estimates in LST 0-8h exceed the SKA1-Low requirements (red crosses in Figure \ref{fig:sensBW}) by factors ranging from $\sim$1.2 (at 70 MHz) to $\sim$2.3 (at 55 MHz). 

\begin{table}[t]
\caption{Estimated SKA1-Low sensitivities (average between XX and YY polarization).} 
\label{tab:sensit}
\begin{center}       
\begin{tabular}{|c|c|c|} 
\hline
\rule[-1ex]{0pt}{3.5ex}  $\nu$  & $<s_{SKA}>_{\text{0-8h}}$  &  Requirement\cite{caiazzo}  \\
\rule[-1ex]{0pt}{3.5ex}   (MHz) & (m$^2$ K$^{-1}$) & (m$^2$ K$^{-1}$)  \\
\hline
\rule[-1ex]{0pt}{3.5ex}  $55$  & $150$ & $64$\tablefootnote{This value, not provided in SKA Requirements Specification \cite{caiazzo}~, have been extrapolated between 50 and 60 MHz from a fifth degree polynomial fit of the requirements data.}  \\
\hline
\rule[-1ex]{0pt}{3.5ex} $70$ & $210$ &  $141$  \\
\hline
\rule[-1ex]{0pt}{3.5ex}  $110$  & $630$ & $531$    \\
\hline
\rule[-1ex]{0pt}{3.5ex}  $160$ & $810$ & $610$    \\
\hline
\rule[-1ex]{0pt}{3.5ex}  $230$  & $700$ & $608$   \\
\hline
\rule[-1ex]{0pt}{3.5ex}  $320$  & $810$ &  $549$   \\
\hline
\end{tabular}
\end{center}
\end{table}


\section{Conclusion and future work}

The work presented in this paper provides an initial characterization of the SKA1-Low prototype station AAVS2 performance such as calibratability and sensitivity. 
We have used commissioning AAVS2 snapshot observations at six different frequencies, from 55 to 320 MHz, selected to sample the SKA1-Low bandwidth. 

We have verified the array calibratability and all-sky imaging capabilities of the station (used as a stand-alone interferometer) with simple Sun-based calibration, obtained using the Sun as a point-like calibration source at its transit and transferring the solutions to 22-24 hours snapshots data collected every 5 minutes (Section \ref{sec:obs}). The achieved good quality of images confirms calibration and system stability over long timescales (Section \ref{sec:obs}). Our initial consistency checks on selected radio sources flux densities also corroborate this finding, as they show reasonably small variations ($\leq$20\%) within $\approx$1 hour across their transit (Section \ref{sec:fluxstab}). Moreover, the quality of  absolute flux scales derived through first-order calibration methods is relatively good, with integrated spectra of Centaurus A, Fornax A and Pictor A following quite well the expected power-laws extrapolated from the literature measurements (Section \ref{sec:fluxstab}).  

We have also derived \textquotedblleft zenith\textquotedblright\, sensitivity estimates through the difference imaging technique (Section \ref{sec:sens}). For this analysis, self calibration during daytime, with elevation of the Sun $\geq$45$^{\circ}$, was also applied (Section \ref{sec:obs}). The comparison between self and single snapshot calibration sensitivities shows that they are consistent with each other at frequencies $\leq$ 160 MHz (see Section \ref{sec:obs}), confirming that the system calibration is stable over several hours. Another important result of this work is that our sensitivity estimates, averaged between the two linear polarizations and in LST range $0-8$~h (where T$_{ant}(\nu)$ is closer to the uniform  T$_{sky}(\nu)$ used to compute the SKA1-Low specifications) are from $\sim$1.2 to $\sim$2.3 times the corresponding SKA1-low requirements. Moreover, they are in good agreement with the EM sensitivity simulations (differences $\lesssim$13\%, Section \ref{sec:sensimul}).  

For future work, we plan to extend this analysis using additional commissioning observations and observing frequencies, both already available and to be performed by the end of 2021 (e.g. as long as AAVS2 station will remain operational).
Different calibration methods for the lowest frequencies ($\leq$70 MHz), such as self calibration in the night-time LST ranges using a set of model bright sources in the observed field of view, or against an all-sky model for the diffuse emission (with the Sun added), will allow to improve the accuracy data calibration, imaging and sensitivity analysis. Moreover, we plan to use our sky temperature simulations (Section \ref{sec:sensimul}) to provide estimates of the SKA1-Low sensitivity as a function of T$_{sky}$ variability. This will be presented in future publications. 

Remarkably, the work  presented here is an important step forward in the SKA-Low project towards construction, as the roll out of the telescope is approaching. The construction of the production prototype Aperture Array 0.5 (AA0.5), consisting of 6 full SKA-Low stations like AAVS2, is expected to start in the early 2023\cite{vanes2020}~. Hence, an analysis similar to the one presented here will be  extended to AA0.5 interferometric observations, to verify its performance.


\subsection* {Acknowledgments}

AAVS2 is hosted by the MWA under an agreement via the MWA External Instruments Policy. This  work makes use of the Murchison Radio-astronomy Observatory, operated by CSIRO. 
We acknowledge the Wajarri Yamatji people as the traditional owners of the Observatory site.


\bibliography{report}   
\bibliographystyle{spiejour}   


\vspace{2ex}\noindent\textbf{Giulia Macario} is a post doctoral fellow at INAF Arcetri Astrophyisical Observatory where, since 2019, she works full time on characterizing AAVS2 through commissioning observations, as member of the INAF SKA-low calibration group. 
She received her PhD in Astrophysics from the University of Bologna, Italy, in 2011. Her main research interest has been related to observational studies of diffuse radio emission in galaxy clusters at low frequencies. She has also been a member of the LOFAR Surveys Key Science Project, contributing to LOFAR commissioning activities.

\vspace{1ex}
\noindent Biographies and photographs of the other authors are not available.

\listoffigures
\listoftables

\end{spacing}
\end{document}